\documentclass[12pt,a4paper]{report}
\usepackage{graphicx}
\usepackage{color}
\usepackage{amssymb}
\usepackage{float}
\usepackage{epsfig}
\usepackage[english]{babel}
\usepackage[utf8x]{inputenc}
\usepackage{sidecap} 
\usepackage[superscript,biblabel]{cite}
\usepackage{amsmath}
\usepackage{multicol}
\usepackage[normalem]{ulem} 
\usepackage{soul} 
\usepackage{wrapfig}
\usepackage{epstopdf}
\usepackage[dvipsnames]{xcolor}
\usepackage{here}
\usepackage{xcolor}
\usepackage{caption}
\usepackage{subcaption}
\usepackage{authblk}
\usepackage{float}
 
\usepackage[margin=1in,footskip=0.25in]{geometry}

\usepackage{psfrag}
\def\beq{\begin{equation}}
\def\eeq{\end{equation}}
\floatstyle{boxed}
\begin{document}
\title
{
\LARGE{Emergence of a weak topological insulator from the Bi$_x$Se$_y$ family and the observation of weak anti-localization}
}

\author[1]{Kunjalata Majhi\footnote{These authors have equally contributed to this work.}}
\author[2]{Koushik Pal$^{\dag}$}
\author[3]{Himanshu Lohani}
\author[1]{Abhishek Banerjee}\author[3]{Pramita Mishra}\author[1,4]{Anil K Yadav}
\author[1]{R Ganesan}\author[3]{BR Sekhar}\author[2]{Umesh V Waghmare}
\author[1,5]{PS Anil Kumar}

\affil[1]{\textit{Indian Institute of Science, Bangalore-560012, India}}
\affil[2]{\textit{Theoretical Sciences Unit, Jawaharlal Nehru Centre for Advanced Scientific Research, Bangalore-560064, India}}
\affil[3]{\textit{Institute of Physics, Bhubaneshwar-751005, India}}
\affil[4]{\textit{Department of Physics, Ch. Charan Singh University, Meerut-250004, India}}
\affil[5]{\textit{Center for Nano Science And Engineering, Bangalore-560012, India}}

\maketitle

\begin{abstract}
	
The discovery of strong topological insulators led to enormous activity  in condensed matter physics and the discovery of new types of topological materials. Bisumth based chalcogenides are exemplary strong three dimensional topological insulators that host an odd number of massless Dirac fermionic states on all surfaces. A departure from this notion is the idea of a weak topological insulator, wherein only certain surface terminations host surface states characterized by an even number of Dirac cones leading to exciting new physics. Experimentally however, weak topological insulators have proven to be elusive. Here, we report a discovery  of a weak topological insulator (WTI), BiSe, of the Bi-chalcogenide family with an indirect  band gap of 42 meV. Its structural unit consists of bismuth bilayer (Bi$_2$), a known quantum spin hall insulator sandwiched between two units of Bi$_2$Se$_3$ which are three dimensional strong topological insulators. Angle resolved photo-emission spectroscopy (ARPES) measurements on cleaved single crystal flakes along with density fucntional theory (DFT) calculations confirm the existence of weak topological insulating state of BiSe. Additionally, we have carried out magneto-transport measurements on single crystal flakes as well as thin films of BiSe, which exhibit clear signatures of weak anti-localization at low temperatures, consistent with the properties of topological insulators.

  \end{abstract}
	
	\section{Introduction}
The discovery of topological insulators(TIs) is one of the recent breakthroughs in condensed matter physics. They have been at the forefront of materials research due to their unique electronic properties and  potential for wide range of applications. Like band insulators, TIs host a bulk band gap and do not conduct electricity in the bulk. However, on the surface they host robust symmetry protected massless metallic states that support spin-polarized conduction\cite{moore,Qi,hasan} \hspace{-2.5mm} . These surface states result from the bulk electronic topology and associated band inversion due to strong spin-orbit coupling. The low energy dynamics of such massless surface states can be described with a Dirac Hamiltonian with spin locked to momentum H=$\hbar v_{F}(\bf{\sigma} \times \bf{k}).\hat{\bf{z}}$, where $\hbar$ is reduced Plank's constant, $v_F$ is the fermi velocity, $\sigma$ stands for Pauli matrices and \textbf{k} is the wave vector. The topology of a time-reversal (TR) symmetry invariant insulator is characterized by  $Z_2$ indices which can be either be zero (topologically trivial) or one (topologically non-trivial)\cite{TI3D,balents}\hspace{-1.9mm}. In 2-D, a single Z$_2$ invariant ($\nu_0$) specifies the topology, where as in 3-D four Z$_2$ invariants are needed ($\nu_0$,$\nu_1$,$\nu_2$,$\nu_3$)\cite{roy,fukane} \hspace{-2mm}. In 3-D, a TR invariant topological insulator can further be classified as strong topological insulator (STI)\cite{zhang,xia,chen} or weak topological insulator (WTI), based on  whether a TI hosts odd or even number of Dirac cones on its surfaces respectively. For an STI, $\nu_0$=1 and ($\nu_1$,$\nu_2$,$\nu_3$) can be 0 or 1, whereas for a WTI, $\nu_0$=0 and atleast one of ($\nu_1$,$\nu_2$,$\nu_3$) should be non-zero\cite{TI3D,balents,roy,kane,fukane} \hspace{-2mm}. While an STI exhibits topological surface states (TSS) with odd number of Dirac cones on any surface\cite{zhang,xia,chen}\hspace{-1.5mm}, a WTI hosts TSS with even number of Dirac cones only at specific surface planes\cite{stern} \hspace{-2mm}. A WTI phase can be adiabatically connected to stacked layers of 2D strong TIs\cite{stern,nmat}\hspace{-1.5mm}, and its surface states are linked to the edge states of each 2D TI layer. Translational-invariant coupling between these layers gaps out the surface states perpendicular to the layer directions\cite{Obuse}\hspace{-1.5mm}, leaving the Dirac cones to manifest only at their side surfaces\cite{stern,takashi,mong} \hspace{-1.5mm}. In this regard, TSSs in STI are always manifested whereas they are either manifested or hidden in case of a WTI \cite{stern,imura,morimoto} \hspace{-1.5mm}.
\\
The surface states of WTI were initially thought to be unstable towards non-magnetic disorder \cite{fukane,nomura,ran,imura,morimoto,yoshimura} \hspace{-1.5mm}. However, it has been shown recently that WTI surface states are robust against strong TR-invariant disorder\cite{morimoto,mong}\hspace{-1.5mm}, similar to those of STIs. These states are further predicted to exhibit intriguing quantum phenomena such as one-dimensional helical modes along dislocation lines \cite{ran,yoshimura} \hspace{-1.5mm}, weak-anti-localization (WAL) effect\cite{mong}\hspace{-2mm}, half quantum spin hall effect\cite{Liu} \hspace{-2mm}. Additionally, WTIs offer new possibilities to obtain 2D TIs by exfoliating a single quantum spin hall (QSH) layer from the bulk which is technically more feasible than fabricating complicated quantum wells\cite{bernevig,konig} \hspace{-2mm}. However, the WTI phase has been proved to be elusive experimentally.  KHgSb\cite{yan} \hspace{-1.5mm}, Bi$_2$TeI\cite{tang} and a superlattice structure of PbTe/SnTe\cite{yang} are theoretically proposed to be  WTIs, but their experimental verifications is yet to be achieved. The bismuth-based  layered compound Bi$_{14}$Rh$_{3}$I$_{9}$\cite{nmat,pouly} is the only known experimental realization of a WTI.
The intuitive design of a WTI material borrows from an idea that a stack of weakly coupled 2D topological insulators may lead to a 3D WTI. For example, Bi$_{14}$Rh$_{3}$I$_{9}$ is essentially a stacked graphene analogue\cite{nmat} \hspace{-2mm}, with each layer exhibiting a large spin orbit interaction at bismuth. Another promising route is to stack alternating bismuth bilayers (2D topological insulator) with a trivial insulator. Bi$_2$TeI follows this paradigm and has been theoretically shown to be a WTI, but with no experimental verification. Primarily, the difficulty in fabrication of these materials in single crystalline bulk or thin film form has proven to be a significant hindrance to research in weak topological insulators.

In this letter, we present possibly the simplest realization of a weak topological insulator, borrowing essentially from two  observations. First, bismuth bilayers are exemplary quantum spin hall insulators\cite{bilayer,kyung,kou,drozdove} \hspace{-1.5mm}. Therefore a stack of coupled Bi$_2$ bilayers should ideally give rise to a 3D WTI. We intuit that stacking of bismuth bilayers sandwiched on either side by a topological insulator instead of a trivial insulator could also possibly host a WTI. Second, and more importantly, it is known that the bismuth chalcogenides exhibit what is known as `infinitely adaptive superlattice' phase \cite{valla} \hspace{-2mm}. The stoichiometry of Bi$_x$Se$_y$ can be adjusted to a wide range of values by composing stacks of Bi$_2$Se$_3$ and Bi$_2$  with thickness of suitable ratio. For the simplest possible ratio with $x=1$ and $y=1$, one obtains a crystal structure as shown in Fig. 1a wherein a repeating unit is formed by a Bi$_2$ bilayer sandwiched bwtween two Bi$_2$Se$_3$ layers. The hallmark of this structure is that it consists of quantum spin hall insulators\cite{bilayer,bi(111),liuZ} that are coupled to each other through 3D topological insulators\cite{zhang,fukane,TI3D,balents} \hspace{-1.5mm}. 

With combined experimental and theoretical studies, we show that BiSe belongs to (0;001) class of $Z_2$ weak TIs, where the band inversion takes place at $\Gamma$ and $A$ points in the Brillouin Zone, which is contributed by both the Bi$_2$ layer and Bi$_2$Se$_3$ quintuple layers. To probe surface transport properties, we grow thin films of BiSe and study  magnetotransport at low temperatures. Magnetoresistance measurements indicate a pronounced weak-antilocalization cusp, consistent with a  strong spin-orbit coupling in the system. 

\section*{Theoretical Analysis}

Electronic structure of BiSe, calculated within a non-relativistic approximation at the experimental lattice constant \cite{Gaudin} (see Fig. 2), reveals a metallic character with a flat conduction band along $\Gamma$-A line just above Fermi level (E$_{F}$), which is highly localized and confined to the bismuth bilayer (see Supplementary Fig. S 1b). Dispersion of this band near the E$_F$ (along $\Gamma$-A) is similar to that of an unconventional superconductor MgB$_2$ \cite{cohen} and weak topological insulators like Bi$_2$TeI \cite{tang} and KHgSb \cite{yan} \hspace{-1.5mm}. This unoccupied conduction band is constituted primarily of  $\sigma$-bonded  p$_x$ and p$_y$ orbitals of Bi in the bismuth bilayer, and becomes dispersion-less due to weak interlayer interaction. Such empty covalent bonds cost energy, and this flat band pushes BiSe to the brink of lattice instability (with  imaginary frequency of $\sim$ $i$15 cm$^{-1}$ appearing along $\Gamma$-A direction, the crystallographic direction of  stacking in BiSe, see Fig. 2d). Orbital projected electronic structure reveals that the contributions to the valence band (VB) and conduction band (CB) around the Fermi level are mainly from the Bi bilayer and Bi$_2$Se$_3$ quintuple layers (QLs) respectively, except along 
the $\Gamma$-A line where this is reversed, indicating the inverted band structure and existence of topologically nontrivial phase in BiSe. With inclusion of spin-orbit coupling in determination of electronic structure ( see Fig. 2b), the doubly degenerate flat bands split opening up a gap along $\Gamma$-A, and elsewhere in the Brillouin Zone (smallest indirect gap being 42 meV). As BiSe is centrosymmetric and invariant under time reversal, the topological Z$_2$ invariant of its electronic structure  effectively determines its topological nature. We determine the four Z$_2$ invariants ($\nu_0; \nu_1, \nu_2, \nu_3$) from the parity of occupied bands at eight time reversal invariant momenta (TRIM) following Fu and Kane's work \cite{fukane} \hspace{-1.5mm}. The strong topological index, $\nu_0$, is calculated by taking the product of the parity  of occupied bands ($\sigma_i$) at eight TRIM through the relation $(-1)^{\nu_0}=\prod_{i=1}^{8}\sigma_i$, where $i$ runs over eight TRIM, and $\sigma_i=\prod_m\xi_{2m}$, $\xi$ being the parity of the 2m$^t$$^h$ occupied band at i-th TRIM. The other three weak topological indices ($\nu_k, k=1,2,3$)  are determined  using $(-1)^{\nu_k}=\prod_{i=1}^4\sigma_i$, where $i$ runs over the four TRIMs (see Supplementary Section IV for detailed analysis). The Z$_2$ invariants of BiSe are (0;001), showing that it belongs to a class of weak topological insulators like KHgSb \cite{yan} and Bi$_2$TeI \cite{tang} \hspace{-1.5mm}.

Goverts et al.\cite{goverts} recently questioned the claim that bismuth bilayer terminated Bi$_2$Se$_3$ exhibits a single  Dirac cone in the electronic structure of (001) surface \cite{he} \hspace{-1.5mm}. They showed that the Dirac-like cone in Bi$_2$ terminated surface  actually corresponds to Rashba split states of the bismuth bilayer which arises from the  presence of an internal electric filed associated with polarity and charge transfer from Bi$_2$ to Bi$_2$Se$_3$ quintuple layers \cite{goverts} \hspace{-1.5mm}. Our results for the electronic structure of (001) surfaces with different surface terminations and thicknesses of BiSe indeed agree with Goverts et al.\cite{goverts} \hspace{-1.5mm}. Thus, the composite system BiSe neither possesses the strong topological properties of Bi$_2$Se$_3$ \cite{zhang} nor of Bi$_2$ layers \cite{wada} (see Supplementary Section V). To further confirm the weak topological nature of BiSe, we determined surface electronic structure on (100) surface (i.e. yz-plane of the unit cell, see Supplementary Section I and Fig. S 3c,d) of BiSe, which reveals two(even number of) Dirac cones at $\bar{Y}$ and $\bar{T}$ points of the Brillouin zone (see Fig. 2c). Similar to  the case of  another weak TI, Bi$_2$TeI \cite{tang} \hspace{-1.5mm}, the two Dirac points ($\bar{Y}$ and $\bar{T}$) are not the projections  of $\Gamma$ and A points of the bulk on the (100) plane of the reciprocal space in Fig. S 3d.

\section*{Angle Resolved Photoemission Spectroscopy}

Angle-resolved photo-emission spectroscopy (ARPES) measurements were carried out on a thin single crystalline flake of BiSe cleaved in situ under a base pressure of 1x10$^{-10}$ mbar. In the ARPES spectrum of BiSe along the $\Gamma$-M direction of the surface Brillouin zone (see Fig. 3a), the high intensity signal around binding energy (BE) E$_{b}$=-0.25 eV corresponds to the bulk valence band (BVB) states, whereas the bulk conduction band (BCB) states are close to fermi energy E$_{F}$ (E$_{b} \simeq -0.03eV$). Interestingly, we find two states in the region between the BCB and BVB showing inverted parabolic dispersion (marked with green dots). These states are marked as surface state bands (SSB), prototypical of Rashba spin-split (RSS) states. The apex of these Rashba Split (RS) states barely touches the E$_{F}$ around K$_{||}$=$\pm$ 0.20 $\AA^{-1}$ (denoted as $k_{0}$). Crossing of these bands (E$_R$) occurs at E$_{b}$ $\sim$ -0.18 eV at the $\Gamma$-point. In the vicinity of this point, these bands exhibit nearly linear dispersion, like Dirac cone states. However, E$_R$ is not clearly visible due to mixing of this feature with the BVB states. The Rashba parameter 2E$_{R}$/k$_{0}$ which has been calculated by using the estimated values of E$_{R}$ and k$_{0}$ from the ARPES data (see Fig. 3a)  is 1.8 eV$\AA$. Comparing these results with the relativistic band structure of BiSe (see Fig. 2b) along the (001) plane, it is clear that the calculated surface states, appearing along the $\Gamma$-M direction (marked by red color box) due to strong spin-orbit coupling (SOC), are in fairly good agreement with the SSB observed in the ARPES image. However, the energy position of the SSB in the calculated band structure is at a lower binding energy than that in the experimental ARPES data. This difference in the energy position could be due to band bending \cite{Ishizaka} and/or due to intrinsic doping from charged defects.
 
In order to understand the isotropic nature of the SSB, we measured the ARPES spectra along two different orientation: $\sim$ $8^{0}$ ($\Gamma$-$M^{'}$) and $\sim$ $15^{0}$ ($\Gamma$-$M^{"}$) offset from the $\Gamma$-M direction. Along the $\Gamma$-$M^{'}$ direction, the second RS state is completely visible(see Fig. 3c), unlike the previous case (see Fig. 3a). However, the separation between the two RS states is not well resolved and a dumbbell shape intensity distribution is observed around the $\Gamma$- point. In the other k-direction $\Gamma$-$M^{"}$, maximum of the second parabolic band appears blurred and near the $E_{F}$ intensity distribution shows a pear shape structure around the $\Gamma$-point, as seen in Fig. 3d. It is clear from these ARPES images that there are two Rashba spin split SSB states in BiSe which arise from the Rashba splitting of states in the Bi bilayer due to potential gradient generated from the charge transfer from Bi bilayer to the Bi$_2$Se$_3$ quintuple layers. Though, the RS states like SSB are not well resolved, these observations are close to our theoretical band-structure prediction of weak topologically insulating phase in BiSe. Secondly, these observations are also consistent with a previous ARPES study on a similar system i.e. Bi films grown on Bi$_{2}$Se$_{3}$\cite{Miao} \hspace{-1.5mm}. On the other hand, these SSB are remarkably different from the SSB found in a typical strong topological insulator, for example Bi$_{2}$Se$_{3}$, shown in Fig. 3e. In this case, the single Dirac cone like feature is clearly visible in the vicinity of E$_F$ near the $\Gamma$-point. This difference in the observed ARPES intensity pattern between BiSe and strong TI, Bi$_2$Se$_3$ further confirms the  topologically distinct character of the SSB in BiSe.

 \section*{Low temperature magnetotransport measurements}

The electronic structure of BiSe has its signatures in the low temperature magneto-transport properties (see Fig. 4a). We find that its resistivity increases with temperature, marking a metallic behavior. The carrier concentration and mobility estimated from Hall effect measurement are  $n_{e}=5.14 \times 10^{20} cm^{-3}$ and  $\mu=722 cm^{2} V^{-1} s^{-1}$ respectively for a single crystal flake of thickness 0.22mm. This carrier concentration can be attributed to selenium vacancy defects as corroborated by our ARPES measurements, but a large bulk carrier concentration proves to be a major obstacle in sieving out transport properties of only the surface states. Resistance measured in a perpendicular magnetic field (see Fig. 4b) produced non-saturating linear magnetoresistance. Though, the origin of linear magnetoresistance is highly debated, we assume this might be due to the continously varrying fluctuations in the conductivity in the weak disorder limit\cite{PL}\hspace{-1.5mm} (discussed later).  

A large bulk conduction observed in single crystal flakes of thickness of the order of millimeters can be quantified with dimensional scaling of properties of films. Hence, we grew thin crystalline films of BiSe using pulsed laser deposition(PLD) on SiO$_2$(500nm)/Si(111) substrates. We used X-ray diffraction and atomic-force microscopy to characterize the crystallinity of our thin films and determine the surface morphology respectively (see Supplementary section I).
Electrical transport measurements were carried out on two BiSe films patterned into standard six probe Hall bars of different thickness, both in the ultra-thin regime. In the zero-field resistivity $\rho_{xx}$ data for 9nm and 15nm samples as a function of temperature (see Fig. 5a), a sharp low-temperature resistivity upturn below $\sim$30K is indicative of an insulating ground state at low temperatures. At T=230K, we observe another transition from metallic to semiconducting state. This can be explained as follows: in the two channel model with a metallic surface in parallel with a semiconducting bulk\cite{pan}\hspace{-1.5mm}, the net conductance of the sample is dominated by the more conducting channel. As a function of increasing temperature, the conductance of the metallic channel decreases linearly while that of the semiconducting bulk increases exponentially. One therefore expects a cross-over temperature above which conductance of the semiconducting bulk exceeds the conductance of the metallic surface states originating from the Rashba-spin split bands, driving a high temperature transition to semiconducting behavior. This is what we observe for the 15nm film. However, for the 9nm film, where the bulk conduction is smaller, such a transition is expected to happen at a higher temperature, and we therefore do not observe it in our measurements till 300K. We observed linear Hall resistance for both 15nm and 9nm samples. The linearity indicates that majority of charge cariers are of simillar type and mobility\cite{bansal}\hspace{-1.5mm}. The carrier concentrations of both the samples are of the same order with $n_{e}=9.3 \times 10^{19} cm^{-3}$ for 15 nm sample and $n_{e}=4.7 \times 10^{19} cm^{-3}$ for 9 nm sample. Mobility of the samples is low: with $\mu=10 cm^{2} V^{-1} s^{-1}$. Such low mobilites indicate strong scattering from static disorder present in the sample. Assuming a spherical Fermi surface\cite{Dieter}, we estimated $\gamma = \frac{1}{\pi k_f l}$, where $\gamma$ is a dimensionless parameter that measures the strength of disorder, k$_f$ is fermi wave vector and $l$ is the mean free path, to be 0.25 and 0.40 for 15nm and 9nm respectively ($\gamma$ $\ll$ 1, corresponds to weak disorder), which indicates that our films fall in the  weak-disorder regime.
In the limit of weak disorder, weak localization corrections to conductivity are expected to appear at low temperatures. By applying a small magnetic field, the weak localization correction to conductivity can be destroyed thereby forming a powerful tool to probe weakly disordered conductors. A wealth of information about electron dynamics including phase relaxation lifetimes, nature of disorder and symmetry classes of governing Hamiltonians can be extracted from these measurements.

Since, 3D topological insulators (both STIs and WTIs) belong to the symplectic class in Altland-Zirnbauer classification\cite{atland} \hspace{-1.5mm}, they exhibit weak anti-localization, with correction to the Drude conductivity arising from destructive interference of time-reversed paths. With  application of magnetic field, time reversed trajectories accumulate opposite phases, eventually leading to decoherence when the phase difference is of order $\sim\pi$. This magnetic field induces suppression of weak anti-localization corrections in two dimensional systems is quantified by the HLN equation \cite{HLN}\hspace{-1.5mm},
$\Delta G_{WAL}(B)=
\Delta G_{xx}(B)-\Delta G_{xx}(0)=\alpha \frac{e^{2}}{2\pi^{2}\hbar}[ln(\frac{\hbar}{4BeL_{\phi}^2})-\Psi(\frac{1}{2}+
\frac{\hbar}{4BeL_{\phi}^2})]$,
where e is charge of electron, $L_{\phi}$ is the phase coherence length, $\alpha$ is WAL co-efficient and $\Psi$ is the digamma function.

 We measured magneto-resistance at different temperatures from 2K to 25K in a perpendicular B-field upto 7T (see Fig. 5b). At low temeperatures, in contrast to single crystal flake, a well-defined weak-anti-localization cusp is clearly seen in the low field region that graduates into a linear magnetoresistance regime for higher fields. Fitting the HLN equation upto 0.5T, we find out the phase coherence $L_{\Phi}$ (45nm, 73nm at 2K) and $\alpha$ (0.33, 0.42 at 2K) for 9nm and 15nm films respectively (see Fig. 5b and 5d). From WAL point of view, $\alpha$ corresponds to the number of 2D coherent conduction channels present in the system. A coherent channel having a $\pi$-Berry's phase should contribute a value of 0.5 to $\alpha$. We observed that the value of $\alpha$ remains close to 0.5 (0.33 and 0.42 for 9m and 15nm respectively) indiacting a single coherent conducting channel contributing to the conductance.  The phase-coherence length increases moderately from 45nm to 73nm with thickness. It is interesting to note that parallel field yeilds significantly lower magnetoresistance compared to perpendicular field configuration simillar to STIs (see Supplementary section II), which needs to be understood further.

 The observed WAL is a result of destructive intereference of electron paths due to $\pi$ Berry's phase in topological insulators. The appearance of WAL in our system indicates the presence of topologically protected surface states, also unlike topologically trvial systems it doesn't exhibit any cross-over from WAL to WL\cite{LiuM} \hspace{-1.5mm}. Spin orbit coupling of the Rashba type leads to chiral fermions with spin locked perpendicular to the direction of momentum. Such states contribute to quantum transport through weak anti-localization and have been observed in several other systems\cite{koga}\hspace{-1.5mm}. Functionalization and control of such states is highly desired, and has received tremendous recent attention as they form building blocks of next generation electronic technologies like Majorana-fermion based quantum computing and spintronics. Transport properties of WTIs may offer exciting new opportunities in these directions, apart from forming a fertile ground for investigating new topological physics. 

\section*{Conclusion}

Weak topological insulators are relatively rare phases of topological matter. While a wealth of information about this phase is available from theoretical analysis, a thorough experimental verification and understanding have been lacking. We demonstrate that BiSe, belonging to the richly explored bismuth chalcogenide family, is a weak topological insulator. With first-principles calculations, we reveal that BiSe exhibits a pair of band inversions at $\Gamma$ and $A$ points in the 3D Brillouin Zone, and calculation of the four $Z_2$ topological indices confirm the WTI phase. Strikingly, we find that bismuth bilayer terminated (001) surface reveals gap-less Rashba spin split states. ARPES measurements on single crystals cleaved in the (001) direction provide clear indications of surface states showing Rashba spin splitting, and other features of the spectra match closely with the calculated electronic sturcture. To investigate the `weakness' of a WTI, we perform electronic transport measurements on BiSe thin films that reveal signatures of weak anti-localization. Thus, we provide a simple route towards fabrication of a WTI amenable to exfoliations and point out the experimental signatures of this 3D topological phase, that will guide developement of other WTI's.

\section*{Ackowledgement}
KM thanks CSIR India for the financial support. PSAK thanks Nanomission, DST, India for financial support. UVW thanks DST, India for support through a JC Bose National fellowship.

\section*{Methods}
\subsection*{Sample preparation and transport measurement}
Single crystals of BiSe have been grown by modified Bridgman method and characterized by Laue method shows good crystalline quality. Thin films of BiSe have been prepared by Pulsed Laser Deposition with laser source of wavelength 248nm and energy 31mJ. To study transport properties we preferred to grow BiSe on Si(111) with SiO$_{2}$(500nm) substrates to avoid any kind of substrate contribution to the sample as the resistivity of the samples are very high. Standard six-probe hall bar shaped thin films were prepared by shadow masking for transport measurements. For electrical contacts, Cr/Au (10nm/65nm) were deposited by thermal evaporation method. The samples were wire-bonded (Al wire) to a PCB. All the transport measurements are carried out in an Oxford 2K system at a magnetic field up to B=7T. 
\subsection*{ARPES measurement}
ARPES spectra were collected using a high flux GAMMADATA VUV He-I (21.2eV) source and SCIENTA R3000 analyzer. At the HeI line (21.2 eV), the photon flux is of the order
of 1016 photons/s/steradian with a beam spot of 2mm. Fermi energies were
calibrated by using a freshly evaporated Ag film on the sample holder. The
total energy resolution estimated from the width of the Fermi edge, was
about 27 meV for HeI excitation energy while the angular resolution better
than 1\AA\hspace{0.5mm} in the wide-angle mode (8\AA) of the analyser.
			\begin{figure}[ht!]
				\centering
				\includegraphics[width=0.95\textwidth]{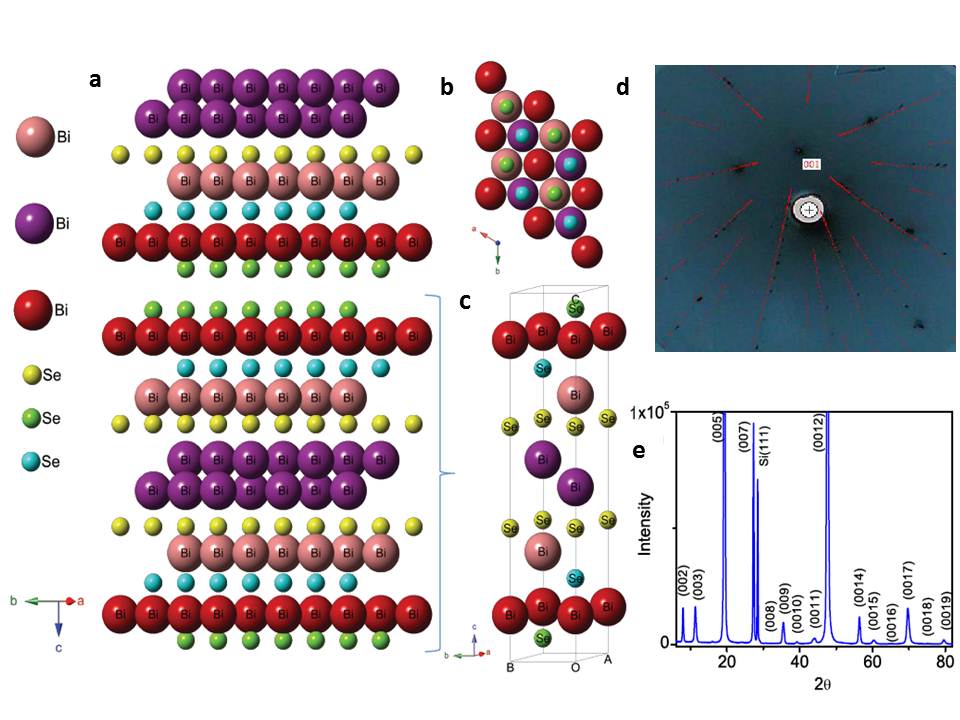}
				\caption{\textbf{Crystal structure and characterization.} \textbf{a}, simulated crystal structure of BiSe, which consists of a bilayer of bismuth sandwiched between two Bi$_{2}$Se$_{3}$ layers. \textbf{b}, top view of the system. \textbf{c}, a single unit cell of BiSe of length 22.9\AA. \textbf{d}, Laue diffraction pattern, revealing the high crystalline quality of grown BiSe single crystals. The red dots are the superimposed simulated Laue pattern. \textbf{e}, X-ray difraction patterns  recorded on a thin film is shown. The observed data illustrates highly c-axis$(000\emph{l})$ oriented growth of BiSe thin films by Pulsed Laser Depsition(PLD).}
			\end{figure}
			\begin{figure}
				\centering
				\includegraphics[width=0.95\textwidth]{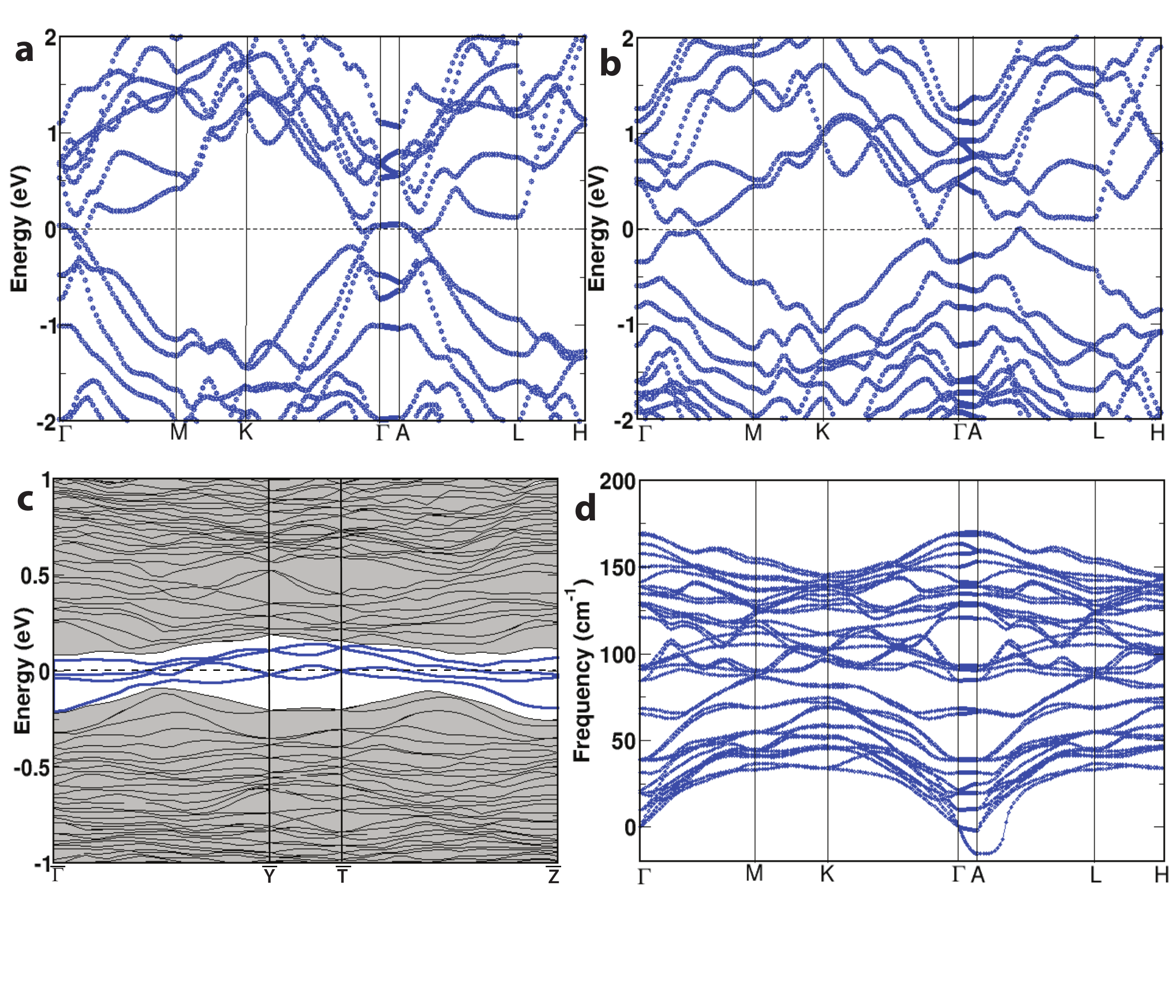}
				\caption{\textbf{Electronic structure and phonon dispersion of BiSe.} {\textbf{a}}, Electronic structure of BiSe calculated without spin-orbit interaction reveals a metallic state. \textbf{b}, spin-orbit interaction opens up a band gap throughout the  Brillouin zone, making BiSe a small  ($\sim$ 42 meV) indirect band gap semiconductor. \textbf{c}, electronic structure of (100) surface of BiSe reveals two Dirac cones(highlighted with blue colors) at $\bar{Y}$ and $\bar{T}$ points in the surface BZ; here, grey shaded regions represent the bands arising from the bulk(see BZ in Supplementary  Fig. S 3d). The empty dispersion-less conduction band along $\Gamma$-A in \textbf{(a)} makes  BiSe  weakly unstable, as evident from its phonon dispersion \textbf{(d)} calculated without including  spin-orbit coupling.}\label{fig:1}
			\end{figure}
				\begin{figure}
					\centering
					\includegraphics[width=0.95\textwidth]{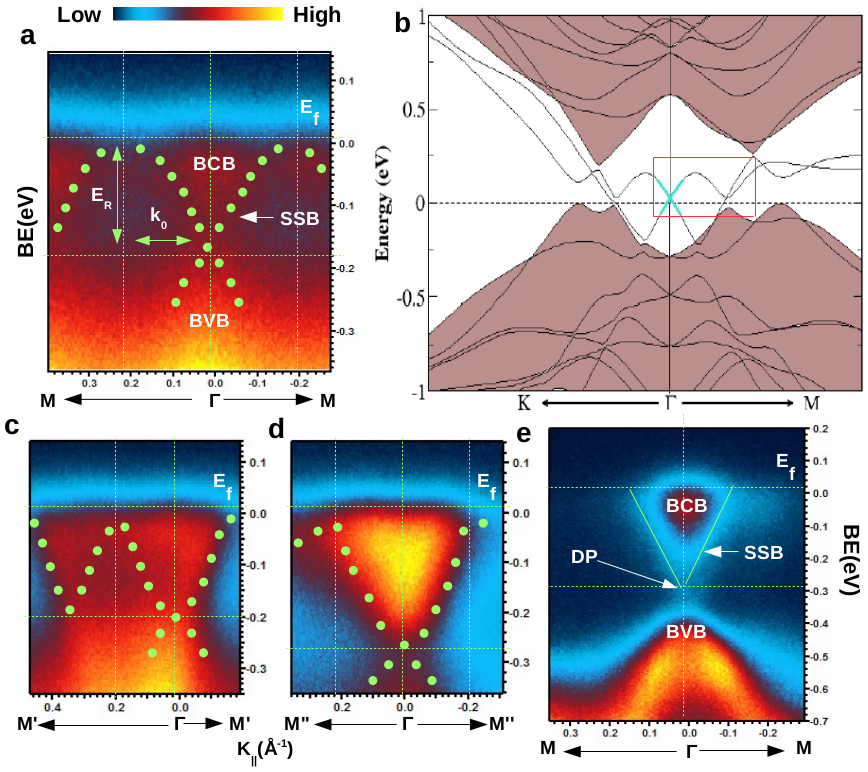}
             	\caption{\textbf{Angle-resolved photoemission spectroscopy on BiSe.} \textbf{a}, the ARPES spectrum along the $\Gamma$-M direction of the surface Brillouin zone of the BiSe. Surface state band (SSB), bulk valence band (BVB) and conduction band (BCB) are marked. \textbf{b}, relativistic surface electronic structure of bismuth bilayer terminated
        		BiSe calculated on (001) plane. Red color box encloses Rashba spin split (RSS)
        		states, where Dirac-like linear dispersing  is highlighted with cyan color. \textbf{c}, \textbf{d}, show ARPES intensity plots taken along $\Gamma$-$M^{'}$ and $\Gamma$-$M^{"}$ directions which are $8^0$ and $15^0$ away from the perfect $\Gamma$-M direction respectively. Difference in the energy position of crossing of RS like bands(E$_R$) between \textbf{(a)} and \textbf{(d)} could be possibly a clear signature of band bending effect because ARPES spectra was taken after half an hour from the cleaving while later case it was collected after four hour from the cleaving. \textbf{e}, displays ARPES intensity plot of Bi$_{2}$Se$_{3}$ along the $\Gamma$-M direction. The bulk and surface states are clearly distinguished and Dirac point (DP) occurs around binding energy $E_{b}$ = -0.28 eV.}\label{ARPES}
			\end{figure}
			
		\begin{figure}
			\centering
			\includegraphics[width=0.95\textwidth]{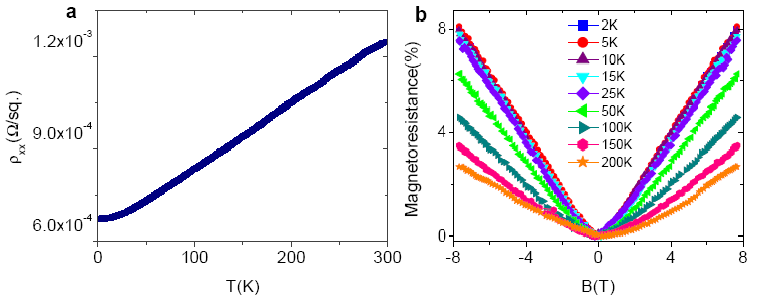}		
			\caption{\textbf{Electrical transport measurement on BiSe single crystal.} {\textbf{a}}, $\rho_{xx}$ as a function of temperature shows linear behavior with temperature of a single crystal flake of thickness 0.22mm and {\textbf{b}}, shows linear magnetoresistance as a function of magnetic field at different temperatures.}
		\end{figure}
		\begin{figure}
		\centering
		\includegraphics[width=0.95\textwidth]{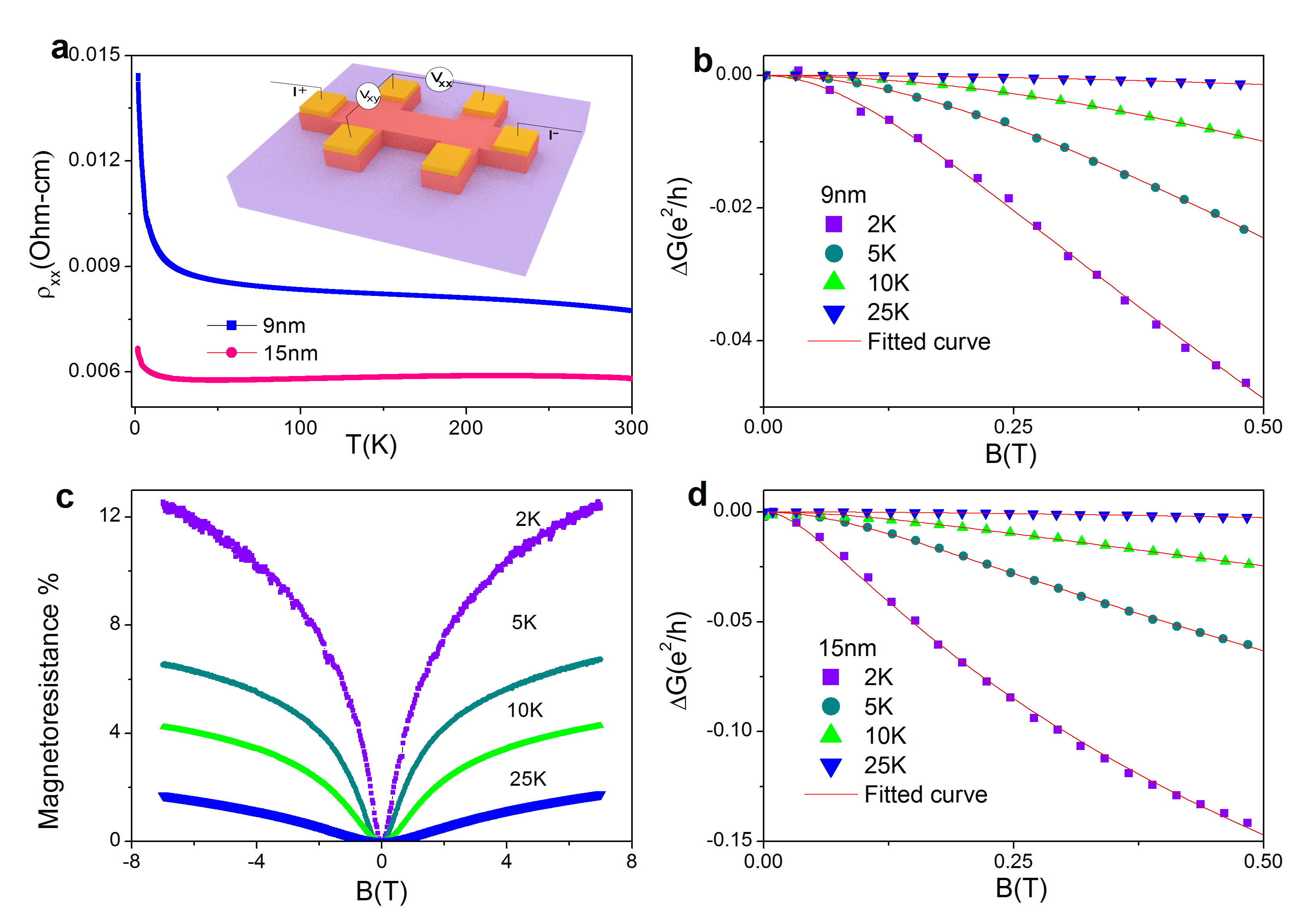}		
		\caption{\textbf{Electrical transport on BiSe thin films.} \textbf{a}, $\rho_{xx}$ as a function of temperature of 9nm and 15nm thin film showing insulating behavior(inset schematic of a Hall bar). \textbf{c}, Magnetoresistance  as a function of applied magnetic field of 9nm film at various temperatures. \textbf{b}, \textbf{d}, HLN fitting for both 9nm and 15nm at different temperatures. \textbf{e}, \textbf{f}, variation of $\alpha$ and L$_{\phi}$ with temperature.}\label{fig:transport}
	\end{figure}

\end{document}